\documentclass[prd,superscriptaddress,11 pt,preprintnumbers]{revtex4}

\usepackage{amsfonts}
\usepackage{amsmath}
\usepackage{amssymb}
\usepackage{bm}
\usepackage{dcolumn}
\usepackage{epsfig}
\usepackage{graphicx}
\usepackage{graphics}
\usepackage[latin1]{inputenc}
\usepackage{latexsym}
\usepackage{rotating}
\usepackage{hyperref}

\newcommand\be{\begin{equation}}
\newcommand\ba{\begin{eqnarray}}
\newcommand\ee{\end{equation}}
\newcommand\ea{\end{eqnarray}}

\begin{document}

\title{Chern-Simons Inflation and Baryogenesis}

\author{Stephon Alexander}
\affiliation{Department of Physics, The Koshland Integrated Natural Science Center, Haverford College
Haverford, PA, 19041} 
\affiliation{Department of Physics, Princeton University, New Jersey 08544, USA}
\affiliation{Department of Physics and Astronomy, Dartmouth College, Hanover, NH 03755}

\author{Antonino Marcian\`o}
\affiliation{Department of Physics, The Koshland Integrated Natural Science Center, Haverford College
Haverford, PA, 19041} 
\affiliation{Department of Physics, Princeton University, New Jersey 08544, USA}
\affiliation{Department of Physics and Astronomy, Dartmouth College, Hanover, NH 03755}

\author{David Spergel}
\affiliation{Department of Astrophysical Science, Princeton University, New Jersey 08544, USA}

\date{\today}

\begin{abstract} 
\noindent 
We present a model of inflation based on the interaction between a homogeneous and isotropic configuration of a $U(1)$ gauge field and fermionic charge density $\mathcal{J}_{0}$. The regulated fermionic charge density is generated from a Bunch-Davies vacuum state using the methods of Koksma and Prokopec \cite{Koksma:2009tc}, and is found to redshift as $1/a(\eta)$.   The time-like component of gauge field is sourced by the fermionic charge leading to a growth in the gauge field $A(\eta)_{0}\sim a(\eta)$.  As a result inflation is dominated by the energy density contained in the gauge field and fermionic charge interaction, $A_{0}\, \mathcal{J}^{0}$, which remains constant during inflation.  We also provide a mechanism to generate a net lepton asymmetry.  The coupling of a pseudo scalar to the Chern-Simons term converts the gauge field fluctuations into lepton number and all three Sahkarov conditions are satisfied during inflation.   Finally, the rapid oscillation of the pseudo scalar field near its minimum thermalizes the gauge field and ends inflation.  We provide the necessary initial condition on the gauge field and fermionic charge to simultaneously generate enough e-folds and baryon asymmetry index.  
\end{abstract}


\maketitle


\section{Introduction}

\noindent The most successful paradigm of the early universe is cosmic inflation.  Despite its usefulness in model building, a number of conceptual and technical challenges associated with inflation driven by fundamental scalar fields have been discussed in the literature \cite{Brand}. Inflationary models driven from vector fields have been considered in the past beginning with the work of Ford \cite{Ford:1989me} followed by other authors \cite{Muk,Jab}.  Building on these past investigations, we present a model of inflation with new ingredients involving the interaction of an abelian gauge field with a fermionic charge, as opposed to solely gauge fields, to generate a realistic inflationary epoch.  
The mechanism that we pave a possible way to generate the net-lepton asymmetry by realizing the Sakharov conditions \cite{Sakharov} during inflation by dynamically intra-converting  the gauge field fluctuations into the lepton asymmetry of the universe, building on previous work of \cite{Alexander:2004us}.  

We now wish to spell out the basic idea of the mechanism before its mathematical instantiation.  We find that the negative pressure equation of state responsible for inflation is generated by a nearly constant interaction energy between the gauge field and fermionic current $V_{int} \sim A_{0}\,  \bar{\psi}\gamma^{0}\psi \, e^{\, 0}_0$ (as opposed to potentials that are purely functionals of gauge fields $V(A_{\mu}A^{\mu})$).   We assume that the early universe is dominated by time-like, homogeneous and isotropic configuration of an abelian gauge field, $A_{0}$, and a ground state of fermion charge density $J^{0}$.  In an expanding space-time, a gauge field and a fermion current typically dilute.  However, we find new solutions of the coupled field equations demonstrating that the gauge fields are amplified due to the quantum effects of the accelerating space-time on the fermionic charge density.  Although the fermion current will dilute as $\mathcal{J}^{0}\!=\!J^{0} e^{\, 0}_0 \sim \frac{1}{a}$ with cosmic expansion, the gauge field amplification will compensate and a constant interaction energy between the fermion and gauge field continues to persist within the horizon scale, which drives inflation. We will also show that the gauge field perturbations are suppressed during inflation. 

A crucial role that the Chern-Simons term plays in this model concerns one-loop quantum effects between gauge fields and fermions in the standard model ({\it i.e.} the Chiral Anomaly).   Consequently,  the Chern-Simons interaction is active during the inflationary epoch so as to naturally satisfy all three Sakharov conditions and can naturally end inflation by the production of an observed baryon asymmetry when the gauge field is converted into leptons through the Chiral-Anomaly \cite{Alexander:2004us}.  Assuming an instantaneous reheating, we calculate the net baryon number to be $\frac{n}{s} \sim 10^{-10}$ in terms of the gauge fields that sourced inflation, making a further connection between gauge field driven inflation and baryogenesis; a relation between the observed baryon asymmetry and the initial density of gauge fields necessary for the onset to inflation.  

\section{The Theory} 

\noindent In this model, we assume that the early universe is dominated by an abelian gauge field interacting with the fermionic current whose symmetry group is $U(1)$.  Specifically we can consider this abelian sector to be identified with the hypercharge gauge group of the standard electroweak theory, before electroweak symmetry breaking, $U(1) \rightarrow U(1)_{Y}$.  In this case we will be able to provide a possible leptogenesis mechanism.  However, it is not necessary that this $U(1)$ be identified with the visible sector, as it can be also a dark copy $U(1)_D$, which is quite ubiquitous in string theory compactifications and supersymmetric extensions to the standard model.  In this case, it is more difficult to instantiate a leptogenesis mechanism.  We discuss these possibilities in Section VI.

For our model, we consider the following action:
\begin{eqnarray} \label{action1}
\!\!\!\!S= S_{D}\!+\!\!\int_{\mathcal{M}_4}\!\!\!\!\!\! d^4x\sqrt{-g} \Bigg[\frac{M_p^2\,R}{8\pi} \!-\! \frac{1}{2}\partial_{\mu}\theta\,\partial^{\mu}\theta \!+\! m^2\,\theta^2- \frac{1}{4} F_{\alpha\beta}F^{\alpha\beta}\!+\!\frac{\theta}{4 M_*} F_{\alpha\beta}\tilde{F}^{\alpha\beta}  \Bigg]\,,
\end{eqnarray}
where $S_{D}$ is the covariant Dirac action
\be
S_{D}= \int_{\mathcal{M}_4} d^4x \sqrt{-g} \left(-i \overline{\psi} \,\slash\!\!\!\!\nabla\psi +c.c.+ M\bar{\psi}\psi + q\, \overline{\psi} \,\gamma^I e_I^{\mu} \psi \,A_{\mu}\right)\,. \nonumber
\ee
 The tensor $F_{\mu \nu}\!=\!\partial_{[\mu}A_{\,\nu]}$ is the field strength tensor\footnote{Our theory is quantum mechanically consistent and anomaly free, since the tree level gauge field-fermion current interacts vectorially.  The structure of the vanishing of the one-loop vertices is discussed in detail by Weinberg \cite{Wein}.  However, if the tree-level gauge interaction is coupled to an axial-vector current then there exist no-gauge invariant regulator to restore the violation of the axial-current.  However, this is not the case in our theory.} of $A_\mu$.  We denote as $q$ a dimensionless coupling constant. The fermionic current is $\mathcal{J}^\mu\!\equiv\! \bar{\psi} \,\gamma^I e_I^{\mu} \psi $, being $\psi$ and $\bar{\psi}$ Dirac spinors and $\gamma^I$ with $I=0,...3$ Dirac matrices. We refer to $M_p$ as the Planck mass, $M_*$ is the mass scale of the pseusoscalar decay constant, and $\theta$ is the pseudo scalar responsible for $CP$ violation.  For the purpose of efficiency of the presentation we will evaluate the dynamics of $\theta$ field in Sec.~\ref{consistent} and show that its energy density is ten orders of magnitude smaller than the gauge-fermion interaction, making it insignificant for driving inflation.  To understand which are the relevant terms necessary to generate inflation, let us first compute the energy momentum tensor of the Lagrangian of eq. (\ref{action1}),
\begin{eqnarray} \label{Lagrangia}
\tilde{\mathcal{L}}= {\rm Tr}\left[-\frac{1}{4}g_{\alpha \gamma} g_{\beta \delta}F^{\gamma\delta}F^{\alpha\beta}+\frac{1}{4}\frac{\theta}{M_*}F_{\alpha\beta}\tilde{F}^{\alpha\beta}+ q\,A_{\mu} 
\, \mathcal{J}^{\mu}\right]. \nonumber
\end{eqnarray}
Using the relation $T^{\mu \nu}=  -\frac{2}{\sqrt{-g}} \frac{\delta \sqrt{-g} \tilde{\mathcal{L}}}{\delta g_{\mu \nu}}$, we find that the energy-momentum tensor is:
\begin{eqnarray} \label{tenso}
&T^{\mu}_{\nu}
&= {\rm Tr} \Big[ -q\,\delta^\mu_\nu A_\rho \mathcal{J}^\rho+F_\alpha^{\,\,\mu} F^{\alpha}_{\,\,\nu} -\frac{1}{4} \delta^\mu_\nu g^{\alpha \rho} g^{\beta \sigma} F_{\alpha \beta} F_{\rho \sigma}- q\,A_{(\nu} \mathcal{J}^{\mu)} \Big].
\end{eqnarray}
In this mechanism, we will show that inflation is driven by the purely time-like components of the gauge field $A_{0}$ and fermionic charge $\mathcal{J}_{0}$.  The condition for the gauge field is similar to scalar field driven inflation, where one assumes that inflation is driven by a spatially homogenous classical part plus quantum perturbations\footnote{From now on, we will be working in conformal coordinates $\{\eta,\vec{x}\}$.}, $\phi(x,\eta) = \phi_{0}(\eta) + \delta\phi(x,\eta)$. In our case
\be
A_{\mu}(\eta,\vec{x}) = A^{(0)}_{\mu}(\eta) + \delta A_{\mu}(\eta,\vec{x})=(A^{(0)}_0(\eta)+\delta A_{0}(\eta,\vec{x}),\,A^{(0)}_{i}(\eta)+\delta A_{i}(\eta,\vec{x}))\,.
\ee
The other main quantity entering our analyses, the vacuum expectation value on the Bunch-Davies vacuum of the fermionic charge $\cal{J_{0}}\rm$, is evaluated in Sec. \ref{ferqua} and is found to be solely a function of time.  Furthermore the spatial components of the fermionic current $J^i = 0$, vanishes due to the trace properties of the gamma matrices.  Solutions for the background fields and their fluctuations are shown in detail in the next section.  We now summarize the validity of our approximations, under which space-time is homogenous and isotropic and its dynamics approaches a quasi de-Sitter phase. 

The background solutions enter the energy-momentum tensor in the term that exhibits a negative pressure equation of state and is responsible for inflation, namely $A\!\cdot\! \mathcal{J}\!\equiv\! A_{\mu}\,\mathcal{J}^{\mu}$. Perturbations enter in the electromagnetic energy-density, $(  \vec{E}^2 +  \vec{B}^2 )/(2 a^4)$, which redshift as $a^{-4}$ since both electric and magnetic field propagate as transversal waves. The two sum up to the isotropic energy density 
\be
T_{00}=\!(  \vec{E}^2 +  \vec{B}^2 )/(2 a^4)+ q\, A\!\cdot\!\mathcal{J} = (\vec{\bar{E}}^2 +  \vec{\bar{B}}^2)/(2 a^4) + q\, \bar{A}_0 J^0\,,
\ee
in which $\vec{\bar{E}}\!=\!\vec{E}(\eta=\eta_0)$, $\vec{\bar{B}}\!=\!\vec{B}(\eta=\eta_0)$, $\bar{A}_0\!=\!A_0(\eta=\eta_0)$ and $J^0\!=\!\mathcal{J}^0(\eta=\eta_0)$ are constant in space and time, and $\eta_0$ is some initial time at the beginning of inflation. The de Sitter phase is reached once we impose that the initial electric and magnetic fields, depending on the perturbation $\delta A_{\mu}(\eta, \vec{x})$, are subdominant to the initial gauge-fermion energy $\vec{\bar{E_0}}^2 + \vec{\bar{B_0}}^2 <\!\!< q \bar{A}_0 J^0$.  Then, during inflation, the $A\!\cdot\!\mathcal{J}$ interaction dominates over the the electromagnetic energy-density and the anisotropic terms of the energy-momentum tensor, due to the exponential growth of the temporal component of the gauge field and slow dilution of the time-like fermionic current $\mathcal{J}^{0}$.  

In particular, the off-diagonal terms in the energy-momentum tensor, $A_{(\nu} \mathcal{J}^{\mu)}$, constitute a negligible perturbation with a proper choice of the initial conditions for the background components $A_i(\eta)$. The remaining anisotropic terms in the energy-momentum tensor, due to the Maxwell tensor $F_{\alpha\mu} F^{\alpha}_{\nu}$, are negligible because the background electric field is suppressed as $a^{-2}$ during inflation.  

We will find that the non-vanishing amplitude for the fluctuations of the background field $\delta A_{\mu}(\eta)$, are propagating waves, which will not spoil isotropy at later times provided that the initial fluctuations satisfy $|\delta A_0| <\!\!< A^{(0)}_0$ and $|\delta A_i| <\!\!< A^{(0)}_0$.  Given these initial conditions, we shall now demonstrate that the coupled field equations indeed yield an inflationary epoch.

\section{Gauge Field Dynamics and Initial Conditions} \label{gauge}

\noindent In what follows, we derive solutions to the field equations for the gauge field coupled to both the metric and the fermionic current.  The key to generating inflation is that the scale factor exhibits inflationary behavior when the gauge field and the fermionic current interaction remain nearly constant during inflation due to backreaction of the gauge field. Varying the action with respect to $A_{\mu}$
\ba \label{gy}
\eta^{\gamma \beta}\partial^{\alpha} \!F_{\alpha \beta}\!+\!\varepsilon^{\gamma \alpha \mu\nu}\,  F_{\mu\nu}\, \partial_{\alpha}\theta/(4 M_*)  \! +\! a^4\,  \mathcal{J}^\gamma\!
\!=\!0\,,
\ea
where $\varepsilon^{\gamma \alpha \mu\nu}$ is the Levi-Civita symbol. Eq. (\ref{gy}) are invariant under gauge transformations $A_\mu \rightarrow A_\mu +\partial_\mu \Lambda$, with $\Lambda$ function of the space-time point, as well as the action (\ref{action1}) is gauge invariant under the same transformations. Homogeneity implies for the background solutions $A^{(0)}_\mu$ that $\Lambda$ is constant, if we demand it to be Fourier transformable, or eventually linear in the conformal time depending on the initial conditions. We can then completely gauge fix $\Lambda$ to be constant\footnote{The gauge invariance of the equation of motion prevents from recovering deviations from de Sitter background that we will be imposing below. Quasi-de Sitter dynamics will be anyway provided by higher order terms in $\epsilon$ in Sec. \ref{ferqua}.} in space-time.

We seek to find self consistent solutions to all the above equations of motion, by working in conformal coordinates and assuming the expansion of the universe to be given by a quasi-de Sitter phase, namely $a(\eta)\!=\!a_0/[1-H(\eta-\eta_0)]$ and $\eta_0$ the initial time. In the following equations, (\ref{Lor})-(\ref{adri}) we use the notation $\!\!\dot{\phantom{a}}\!\!=d/d\eta$, while $\nabla_i=\partial/\partial x^i$, and $\mathcal{J}^{i}$ stands for the spatial part of the fermionic current while $\mathcal{J}^{0}$ stands for the temporal part of the fermionic current.

As stated in the preceding section, inflation begins with a time-dependent homogeneous background gauge field, $A^{(0)}_{\mu}(\eta) = (A^{(0)}_0(\eta),\,A^{(0)}_{i}(\eta))$. Moreover, the total gauge potential, including perturbations is
\be
A_{\mu}(\eta,\vec{x}) = A^{(0)}_{\mu}(\eta) + \delta A_{\mu}(\eta,\vec{x})=(A^{(0)}_0(\eta)+\delta A_{0}(\eta,\vec{x}),\,A^{(0)}_{i}(\eta)+\delta A_{i}(\eta,\vec{x}))\,.
\ee
Furthermore, we impose the Lorentz gauge, $\partial_\mu A^\mu$: 
\be \label{Lor}
\dot{A}_0(\eta, \vec{x})\, + \nabla \cdot \vec{A}(\eta,\vec{x})=0\,.
\ee 
The $0$th component of (\ref{gy}) then gives for the temporal background component  
\be  \label{A00}
\ddot{A}^{(0)}_0(\eta) = a^4 \mathcal{J}_0\,,
\ee
in which we have used (\ref{Lor}). We can find a solution for the gauge field provided that $\mathcal{J}_0\sim J_0/a(\eta)$, which we will show to be the case in Sec \ref{ferqua} by computing the expectation value of the fermionic current in the Bunch-Davies vacuum state.  We find that the time-like gauge field is: $A^{(0)}_0(\eta)= \bar{A}_0\, a(\eta)/a_0$, with $a_0\!=\!a(\eta\!=\!\eta_0)$ and $A^{(0)}_0(\eta\!=\!\eta_0)\!=\!\bar{A}_0$ constant in space and time, and $\eta_0$ some initial time for inflation. For the spatial components of the background gauge field $\vec{A}^{(0)}$ the equation of motion is
\be \label{adri}
\vec{\ddot{A}}^{(0)}(\eta)=a^4 \vec{\mathcal{J}}\,.
\ee
Since we will find in Sec. \ref{ferqua} that $\vec{\mathcal{J}}=0$, this latter gives $ \vec{\ddot{A}}^{(0)}=0$, which decays $\vec{A}^{(0)} \simeq \vec{c} + \vec{c}^{\,'}/a(\eta)$.  We assume that there is no background electric field filling the Universe, namely $\vec{c}=\vec{c}^{\,'}=0$.   Nonetheless, isotropy would be preserved even for non vanishing background electric fields, provided that both $|\vec{c}|<\!\!< A^{(0)}_0$ and $|\vec{c}^{\,'}|<\!\!<A^{(0)}_0$.

The equation of motion for the fluctuations around the background field components $A^{(0)}_\mu$, which we assume to be of infinitesimal order in some parameter $\lambda$, are now recovered to be from the variation of (\ref{gy})
\be \label{pert0}
\delta \ddot{A}_0(\eta, \vec{x})\, - \vec{\nabla}^2 \delta A_0(\eta,\vec{x})=0
\,,
\ee
for the $\delta A_0(\eta, \vec{x})$ component, whose solution is trivially found to be 
\be \label{solpert0}
\delta A_0(\eta, \vec{x})=\delta A_0 \, \exp (i k_0 \eta) \, \exp (-i \vec{k} \vec{x})\,.
\ee 
For $\delta \vec{A}(\eta, \vec{x})$ we find
\be \label{perti}
\delta\vec{\ddot{A}}-\nabla^2\delta\vec{A}+\vec{\nabla}\times{\delta \vec{A}}\,\, \dot{\theta}/M_* =0 \,.
\ee
Without loss of generality, we can write the solution of (\ref{perti}) using circular polarization vector fields, setting $\delta A_3$ to be vanishing and assuming the perturbed field to be divergence-less. We can then cast the field equations in terms of Fourier modes 
\be
\delta \vec{A}(x,\eta) = \rm \int d^{3}k \,\sum_{h} \delta A(\eta,k)_{h}\, \epsilon_{h}(k) e^{ikx}\,, 
\ee
where $h=\pm 1$ denotes the two possible helicities. The requirement that $\delta\vec{A}$ is traceless also ensures that $\vec{A}(\eta,k)$ is perpendicular to its direction of propagation. Our field equation for the gauge field then simplifies to
\be \label{fluxequation} 
\delta \ddot{A}(\eta,k)_h + k^2 \delta A(\eta,k)_h= -h \, k\, \delta A(\eta,k)_h \dot{\theta}/M_*\,. 
\ee
Within the assumption of $\dot{\theta}\sim {\rm const}$, which will be justified in Sec.~\ref{consistent}, the general solution for the left-handed gauge field 
is found to be 
\be \label{evoluzione}
\delta A(\eta,k)_{-}=A_-^0\cosh(\beta_k \eta)+\tilde{A}_-^0\sinh(\beta_k \eta)\,,
\ee
where $A_-^0$ and $\tilde{A}_-^0$ are determined by the initial conditions and the growth factor is $ \beta_k^2=k({\dot{\theta}}/{M_*}-k)$.  We will have exponentially growing fields (in conformal time) provided that $\beta_k$ is real  (or $k<\dot{\theta}/M_*$). The gauge field, $A(\eta,k)_{+}$, of opposite helicity also has an oscillatory/exponential behavior, {\it i.e.}
\ba \label{Aminus}
\delta A(\eta,k)_{+}=A_+^0 \cos( \gamma_k \eta) + \tilde{A}_+^0 \sin(\gamma_k \eta)\,, 
\ea
in which $\gamma_k^2=k({\dot{\theta}}/{M_*}+k)$. 

However, if the pseudo-scalar field is slowly rolling during inflation, as considered in \cite{Prokopec:2001nc}, we would have recovered $\dot{\theta}= \dot{\theta}_0/\eta$, with $\dot{\theta}_0$ constant. Accordingly, the equation for the fluctuations of the background gauge potential would have been a Bessel equation \cite{Prokopec:2001nc} (see also \cite{AS} and references therein)
\be \label{ser}
\delta\ddot{A} (\eta, \vec{k})_h + \left[ k^2 + 2 \, h \, \frac{\xi}{\eta}   \right] \delta A (\eta, \vec{k})_h =0\,,
\ee
in which we have defined $\xi= \dot{\theta}_0 /(2 M_*)$. We emphasize that birefringence will still persist in this case, and that solutions, expressed in terms of Bessel functions, may lead to interesting observational parity violating CMB effects. In the perspective of \cite{AS, Sorbo:2011rz}, dissipative effects due to the Chern-Simons coupling can slow down the axion field, leading to inflation even for a steep pseudo-scalar potential \cite{AS}. The eventual detection of parity-violating CMB correlation functions, and precisely non-vanishing TB and EB correlation functions, has been analyzed in \cite{ Sorbo:2011rz}. 

\ In the next section we recover the expression for the components of the fermionic current. 

\section{Fermionic Dynamics} \label{ferqua}

In the previous section we saw that if the time-like fermion current density $\cal{J}\rm_{0}$ scales as $\frac{1}{a}$, the gauge field will grow as $A_{0} \sim a(\eta)$ according to eq (\ref{A00}).  In this section we calculate the vacuum expectation value of the fermion current density and show that in de-Sitter space the physical time-like current indeed scales as $\frac{1}{a}$, while the space-like current vanishes.

\noindent  Let us start by writing the free action for the fermionic field
\be
S=\int d^4 x \sqrt{-g} \left\{ \frac{i}{2} \left[ \,  \overline{\psi} \gamma^\mu \nabla_\mu \psi - \left( \nabla_\mu \overline{\psi} \right) \gamma^\mu \psi \right]  -m \overline{\psi} \psi   \right\}\,,
\ee
where $\gamma^\mu=\gamma^I\, e_I^\mu$, being $\gamma^I$ the Dirac matrices (let say in the chiral representation) and $e_I^\mu$ the inverse of the tetrad defined by $g_{\mu \nu} = e^I_\mu \, e^I_\nu$ (notice indeed that $\{ \gamma^\mu,\, \gamma^\nu \}=-2 g^{\mu \nu}$), and $\overline{\psi} = \psi^\dagger \gamma^0$. The equation of motion derived for $\psi$ are then
\be
i \gamma^\mu \nabla_\mu \psi(x) - m \, \psi(x) = 0\,,
\ee
in which the covariant derivative acting on $\psi$ is expressed in terms of the spin connection
\be
\nabla_\mu \psi(x) = \partial_\mu \psi(x) - \Gamma_\mu \psi (x)\,,
\ee
which in turn may be written in terns of the tetrad and the Christoffel connection as
\be
\Gamma_\mu = - \frac{1}{8} e^\nu_I \left( \partial_\mu e_{\nu J} - \Gamma^\alpha_{\mu \nu } \, e_{\alpha J} \right) \, \left[ \gamma^I, \gamma^J  \right]  
\ee
and is such that $\nabla_\mu \gamma_\nu=0$. In particular, for FLRW metric defined in conformal coordinates by $e^I_\mu=a(\eta) \, \delta^I_\mu$, the equation of motion for $\psi(x)$ is found to be
\be
i \gamma^\mu \, \nabla_\mu \psi(x) = \frac{i}{a^{\frac{5}{2}} (\eta)}  \gamma^I\, \partial_I  \left( a^{\frac{3}{2}} \, \psi(x) \right) = m \psi(x)\,.
\ee 
The Feynman propagator $i S^{ab}_F (x, y)$ of the theory, satisfying at tree level the equation 
\be
\sqrt{-g} \left( i \gamma^\mu \nabla_\mu -m \right)_x i S^{ab}_F (x, y)= \ i \delta^4 (x-y)\, 1\!\!1^{ab} \,,
\ee
is then defined by the relation
\ba
i S^{ab}_F (x, y) &=& \langle  0 |  T \{  \widehat{\psi}^a (x) \widehat{\overline{\psi}}^b (y) \} |  0 \rangle = \nonumber\\
&=& \theta (\eta_x - \eta_y) \  \langle  0 |   \widehat{\psi}^a (x) \widehat{\overline{\psi}}^b (y)  |  0 \rangle -
\theta ( \eta_y - \eta_x) \  \langle  0 |  \widehat{\overline{\psi}}^b (y)   \widehat{\psi}^a (x) |  0 \rangle
\ea
where $a$ and $b$ label spinor indices of the Dirac fields $\widehat{\psi}^a (x)$. 
Notice now that rescaling the fermionic spinor as $\chi(x)= a^{\frac{3}{2}} (\eta) \psi(x)$ we may simplify the equation of motion to be written only in terms of partial derivatives
\be
i \gamma^I \partial_I \chi(x) - a(\eta) m \chi(x) = 0\,.
\ee
Variation of the density Lagrangian with respect to $\dot{\psi}(x)$ provides the conjugate momentum to $\widehat{\psi}(x)$. The two operators undergo second quantization through and are subjected to the anti-commutation relations that are fundamental to the Dirac quantum fields
\be \label{comfer}
\{ \widehat{\psi}_a(\vec{x},t ), a^3(t) \, \widehat{\psi}^*_b(\vec{y},t) \}= i \delta_{a b}
 \delta^{3}(\vec{x} - \vec{y})\,,
\ee
imposed on space-like surfaces. It follows that
\be \label{acr}
\{ \widehat{\chi}_a(\vec{x},t ), \, \widehat{\chi}^*_b(\vec{y},t) \}= i \delta_{a b}  \delta^{3}(\vec{x} - \vec{y})\,, \qquad \{ \widehat{\chi}_a(\vec{x},t ), \, \widehat{\chi}_b(\vec{y},t) \}= 0\,, \qquad \{ \widehat{\chi}_a(\vec{x},t ), \, \widehat{\chi}^*_b(\vec{y},t) \}=0\,.
\ee
Quantized fields may be expanded as
\ba
&& \widehat{\chi}(x) = \int \frac{d^{3} \vec{k} }{(2 \pi)^3} \sum_r \hat{a}_{\vec{k},\, r} \, u_{r} (\vec{k}, \, \eta) e^{i \vec{k} \cdot \vec{x}}  +  \hat{b}^\dagger_{\vec{k},\, r} \, v_{r} (\vec{k}, \, \eta) e^{-i \vec{k} \cdot \vec{x}} \,, \nonumber\\
&& \widehat{\overline{\chi}}(x) = \int \frac{d^{3} \vec{k} }{(2 \pi)^3} \sum_r \hat{a}^\dagger_{\vec{k},\, r} \, \overline{u}_{r} (\vec{k}, \, \eta) e^{-i \vec{k} \cdot \vec{x}}  +  \hat{b}_{\vec{k},\, r} \, \overline{v}_{r} (\vec{k}, \, \eta) e^{i \vec{k} \cdot \vec{x}}\,,
\ea
where $u_r(\vec{k},\, \eta)$ and $v_r(\vec{k},\, \eta)$ are four-spinors labeled with respect to their conformal spatial momentum $\vec{k}$ and the helicity, namely the projection of the spin in the direction of motion, $r=\pm1$. The anti-commutation rules (\ref{acr}) imply that 
\be \label{acro}
\{ \hat{a}_{\vec{k},\, r}, \,  \hat{a}^\dagger_{\vec{k}',\, r'}  \} = (2\pi)^3 \delta^{3} (\vec{k} - \vec{k}') \delta_{r,r'}\,, \qquad \qquad \{ \hat{b}_{\vec{k},\, r}, \,  \hat{b}^\dagger_{\vec{k}',\, r'}  \} = (2\pi)^3 \delta^{3} (\vec{k} - \vec{k}') \delta_{r,r'}\,.
\ee
The fermionic Fock space can be then defined through the action of Ladder operators $a_{\vec{k},\,r}^\dagger$ and $b_{\vec{k},\,r}^\dagger$, which obey the Fermi-Dirac statistics as derived from (\ref{comfer}) and (\ref{acro}), on the vacuum state $|0\rangle$ of the theory. This latter in turn is defined by the action of the Ladder operators $a_{\vec{k},\,r} |0\rangle=b_{\vec{k},\,r}|0\rangle=0$ and corresponds to the Bunch-Davies vacuum state. 

Four-spinors may be further decomposed into a direct product of two-spinors (see {\it e.g.} \cite{Garbrecht:2006jm, Koksma:2009tc}) finding 
\be
u(\vec{k},\, \eta) = \sum_r u_r (\vec{k},\, \eta) = \sum_r 
\Big(\begin{array}{c} 
u_{L,\,r} (\vec{k},\, \eta) \\
 u_{R,\,r} (\vec{k},\, \eta)
\end{array}\Big) \otimes \xi_r\,,
\ee
in which $u_{L,\,r} (\vec{k},\, \eta)$ and $u_{R,\,r} (\vec{k},\, \eta)$ are left-handed and right-handed one-spinors with helicity $r$, and $\xi_r$ is the helicity two-eigenspinor defined by $\widehat{\vec{k}} \cdot \vec{\sigma} \,\xi_r = r \,\xi_r$. In a similar way
\be
v(\vec{k},\, \eta) = \sum_r v_r (\vec{k},\, \eta) = \sum_r 
\Big(\begin{array}{c} 
v_{R,\,r} (\vec{k},\, \eta) \\
 v_{L,\,r} (\vec{k},\, \eta)
\end{array}\Big) \otimes \xi_r\,,
\ee
in which now $v_{R,\,r} (\vec{k},\, \eta)$ and $v_{L,\,r} (\vec{k},\, \eta)$ are right-handed and left-handed one-spinors with helicity $r$. The anti-commutation rules (\ref{acr}) and (\ref{acro}) also imply the relation
\be
\sum_r \ u_r^a (\vec{k},\, \eta) u_r^{*\, a} (\vec{k},\, \eta) +  v_r^a (-\vec{k},\, \eta) v_r^{*\, a} (-\vec{k},\, \eta)= \delta^{ab}\,,
\ee
which we further impose to be subjected to $\overline{u}_{r} (\vec{k},\, \eta) v_{r'} (\vec{k},\, \eta) = 0 = \overline{v}_{r'} (\vec{k},\, \eta) v_{r} (\vec{k},\, \eta)$, following the normalization required in \cite{Garbrecht:2006jm, Koksma:2009tc}. 

Solving the Dirac equation for $f_{\pm r} (\vec{k},\, \eta)\equiv [u_{L,\,r} (\vec{k},\, \eta) + u_{R,\,r} (\vec{k},\, \eta) ]/\sqrt{2}$ for the de Sitter background, we find \cite{Garbrecht:2006jm, Koksma:2009tc} for a mode $\vec{k}= k \hat{{\it k}}$
\be \label{f}
f_{\pm r} (- k \eta) = \pm \, e^{i \frac{\pi }{2} (\nu_\pm + \frac{1}{2})} \sqrt{ - \frac{\pi k \eta}{ 4}} H^{(1)}_{\nu_\pm} (- k \eta)\,,
\ee
where $H^{(1)}_{\nu_\pm}$ are Hankel function of the first kind of order $\nu_\pm = \frac{1}{2} \mp i \zeta $, with $\zeta=m/H$. Identical solutions to $f_{\pm r} (\vec{k},\, \eta)$ in (\ref{f}) are found for $g_{\pm r} (k,\, \eta)\equiv [v_{L,\,r} (k,\, \eta) + v_{R,\,r} (k,\, \eta) ]/\sqrt{2}$. 
Using these results, it is straightforward to find \cite{Garbrecht:2006jm, Koksma:2009tc} 
\ba \label{fp}
&&i S^{ab} (x, y) = a(\eta_x) ( i \gamma^\mu \nabla_\mu + m) \frac{H^2}{\sqrt{a(\eta_x) a(\eta_y) }}\, \left[ i S_+(x,\,y) \frac{1+\gamma^0}{2} + i S_-(x,\,y) \frac{1-\gamma^0}{2}       \right]   \,, \nonumber\\
&& i S_\pm(x,\,y) = \frac{\Gamma(1\mp i \frac{m}{H} ) \, \Gamma(2 \pm i \frac{m}{H} )}{(4\pi)^2 \Gamma(2)}\!\! \phantom{a}_2 F_1\!\left( 1 \mp i \frac{m}{H}, \, 2\pm i  \frac{m}{H}, \, 2, 1-  \frac{\Delta(x,y)}{4} \right)\,, \nonumber\\
&& \Delta(x,y)= a(x) a(y) H^2 \Delta x^2\,, \qquad \Delta x^2= (|\eta_x -\eta_y| - i \varepsilon)^2 + (\vec{x} - \vec{y}) \cdot (\vec{x} - \vec{y}), 
\ea
in which $a(\eta)$ is the de Sitter conformal factor and which has been generalized in \cite{Koksma:2009tc} to the FLRW background, and in which $\varepsilon$ denotes an infinitesimal displacement on the imaginary time line. We then consider\footnote{We acknowledge our referee for pointing out this cut-off independent way of obtaining the expectation value of the fermionic current on the Bunch-Davies vacuum.} that the amplitude of the vacuum expectation value of the fermionic current components may be easily derived from the advanced or retarded propagator entering the definition of the Feynman propagator (\ref{fp})
\be
\langle  0 |   J^I | 0 \rangle \simeq  \ {\rm lim}_{y \rightarrow x } \ S^{ab} (x, \, y)\gamma^I_{ba} 
\,.
\ee
Taking the limit in which the two space-time points coincide, namely $y\rightarrow x $, and the limit in which the infinitesimal displacement shrinks to zero, {\it i.e.} $\varepsilon \rightarrow 0$, we easily recover 
\be \label{res}
S^{ab} (x, y)\simeq \frac{H^2}{16\pi^2}\left[  \frac{m^2}{H} \left( \frac{1+\gamma^0}{2}  \right)^{ab} - \frac{m^2}{H} \left(   \frac{1-\gamma^0}{2}  \right)^{ab}  \right]   +  O\left(\frac{m^2}{H^2}\right) \,,
\ee
in which we have used metric compatibility and expanded the hypergeometric function. 

\ We may now focus on the components of the fermionic current, that we select by tracing the above result (\ref{res}) with the $\gamma^I$ matrices.  We then get that only the temporal component of the fermonic current is non-vanishing and that it can be connected to the relevant results concerning the fermionic propagators in de Sitter space times \cite{Koksma:2009tc}, for fermions with a bare mass $m$ acting as a regulator. The main result of this section then turns out to be
\be
J^0\simeq  \frac{1}{4 \pi^2}\, m^2 H + O\left(\frac{m^3}{H^3}\right) \,,
\ee
and finally for the components of the fermion current $\mathcal{J}^\mu$ 
\be
\mathcal{J}^0=\frac{J^0}{a(\eta)}\simeq  \frac{1}{ 4 \pi^2} \, \frac{m^2 H}{a(\eta)} + O\left(\frac{m^3}{H^3}\right) \,, \qquad  \qquad  \qquad \mathcal{J}^i=0\,.
\ee
We assume that quantum fluctuations of the fields do not spoil the vanishing of the current, consistently with the homogeneity and isotropy of the background de Sitter space during inflation. 
Concluding this section, we address the consistency of the FLRW solutions with the one loop background fluctuations of the matter fields, and study the self consistent semiclassical solutions for the gravitational field.

\section{Consistent Inflationary Dymamics}   \label{ID}

\noindent We now turn our attention to the Einstein Equations and seek an inflationary solution.  The $G_{00}$ component of the Einstein Equations gives the first Friedmann equation:
\ba \label{Fried}
3\frac{\dot{a}^2}{a^4}&=&\frac{8\pi G}{a^4} (\vec{E}^{2} + \vec{B}^2) + 8\pi G \, A_0\,\mathcal{J}^0\,.
\ea
The coupled system can be solved if the interaction term $A\!\cdot\!\mathcal{J}$ is nearly spatio-temporarily constant during inflation.  The fact that $\mathcal{J}_0$ dilutes as $1/a$ might raise concern, but notice that the temporal gauge field component grows proportional to $a(\eta)$, as is evident from solving (\ref{A00}).  Since $A\cdot\mathcal{J}$ is constant we can solve the Friedmann equation (\ref{Fried}) subject to to our gauge field configuration $A_{\mu}(\eta,\vec{x}) = A^{(0)}_{\mu}(\eta) + \delta A_{\mu}(\eta,\vec{x})=(A^{(0)}_0(\eta)+\delta A_0(\eta, \vec{x}),\,A^{(0)}_{i}(\eta)+\delta A_{i}(\eta,\vec{x}))$, choosing for simplicity the initial conditions the electric and magnetic field to be $\vec{E}_{0}(\eta_{0})=\vec{B}_{0}(\eta_{0})=0$.  Therefore the Friedman equations at early times is
\ba \label{Friedsec}
3\frac{\dot{a}^2}{a^4}&=  8\pi G \, A_0\,\mathcal{J}^0\,,
\ea
We then find an inflationary scale factor,
\be \label{dS}
a(\eta)=a_0\, [1- H (\eta-\eta_0) ]^{-1}\,,
\ee
where $a_0$ is a normalization factor such that $a(\eta_0)=a_0$, and  the Hubble parameter is solved to be $ M_p H\simeq \sqrt{(A \!\cdot\! \mathcal{J})_{\eta_0} }a_0 $. We can easily obtain the comoving scale factor $a(t)$ using the map $\partial\eta =\partial t/a$ (namely $a(t)=a_0\exp Ht$), which gives rise to exponential growth in the scale factor.  With the choice of initial conditions such that $a(\eta)/a_0\!=\!(1-H\eta)^{-1}\!=\!\exp Ht$, the time coordinate is conformally mapped to the bounded range $\eta\in[0,H^{-1})$.

As established in Sec.~\ref{gauge} and Sec.~\ref{ferqua}, the equations of motion for the background gauge and fermion charge is constant and we have an inflationary solution.  We have demonstrated that the solution of the gauge field in a time dependent background leads to an energy-density $A\!\cdot\! \mathcal{J}$ that is constant in the Hubble radius at the beginning of inflation. The inflationary solution is self consistent for an initial density of fermion current and gauge field that are constant in a region inside the initial Hubble radius.  The time dependent dynamics of the gauge and fermion field conspire to keep $A\!\cdot\! \mathcal{J}$ constant during inflation. 

We may now ask which kind of evolution we would derive from considering a quasi de Sitter space-time, and whether the system would be provided with consistent solutions even in this case. If we assume a non vanishing, constant and arbitrary $\epsilon=-\dot{H}(\eta)/H(\eta)^2$, then the conformal factor $a(\eta)$ of the FLRW metric is expressed by the relation
\be \label{o}
a_\epsilon(\eta)= a_0 \, \left[1+ (\epsilon-1) H_0 (\eta-\eta_0) \right]^{- \frac{1}{1-\epsilon}} \,,
\ee
in which from now on $H_0$ stands for the time independent Hubble parameter in the idealized de Sitter space-time phase. We then find, using the results of Ref. \cite{Koksma:2009tc}, that 
\be \label{JFLRW}
\mathcal{J}_\epsilon^0=\frac{J^0_\epsilon}{a_\epsilon(\eta)}\simeq  \frac{1}{ 4 \pi^2} \, \frac{m^2 H_0}{(1-\epsilon)a_\epsilon(\eta)} + O\left(\frac{m^3}{H_0^3}\right) \,, \qquad  \qquad  \qquad \mathcal{J}^i=0\,.
\ee
If we expand for $\epsilon <\!\!<1$, which corresponds to a small deviation from pure de Sitter space-time, we obtain the modified fermionic current
\be
\mathcal{J}_\epsilon^0 \simeq \frac{1}{4 \pi^2 } \, \frac{m^2 H_0}{ a(\eta)} \, \left(  1+  \epsilon \ln (1-H_0(\eta-\eta_0) ) \right) + O(\epsilon^2) \,.
\ee
in which $a(\eta)$ corresponds to the scale factor of the de Sitter phase in (\ref{dS}). We then solve consistently for the vector fields and the gravitational fields: 
\be \label{Apsi}
A_0^{\epsilon}(\eta)= \frac{J^0}{2 H_0^2} \, a(\eta)\,
\left( 1- \frac{1}{2}\, \epsilon - 3  \epsilon \ln (1-H_0(\eta-\eta_0) ) \right)\,.
\ee
Together with (\ref{JFLRW}) expanded at first order in $\epsilon$, the expression (\ref{Apsi}) provides for the first Friedmann equation, {\it i.e.} 
\be \label{qdSa}
M_p^2\, H^2_\epsilon(\eta)= \frac{8\pi}{3}\, A^\epsilon_0 (\eta) \, \cal{J}^0_\epsilon(\eta) 
\ee
with $H_\epsilon(\eta)=\dot{a}_\epsilon(\eta)/a_\epsilon^2(\eta)$, the solution 
\be \label{qdS}
a_\epsilon^{(1)}(\eta) = a(\eta)\,\left[1 - \epsilon  \ln \left( 1- H_0(\eta -\eta_0)\right)  \right]\,,   \qquad H_0=\frac{1}{2 \sqrt{3\, \pi^3}} \, \frac{m^2}{M_p}\,.
\ee
Eq. (\ref{qdS}) provides in turn the expansion in $\epsilon$ of eq. (\ref{o}). We have then recovered that a quasi de Sitter solution, consisting of a small and constant $\epsilon$, is consistent with one loop background fluctuations  fermionic fields interacting with the gauge fields. Eqs. (\ref{qdSa}) and (\ref{qdS}) represent non-trivial self-consistent solutions to semiclassical gravity, in which one loop fluctuations causes classical expansion.

\section{Inflationary Leptogenesis} 

\noindent In this section we initiate a potential leptogenesis mechanism during inflation following the logic of \cite{Alexander:2004us}.  As a caveat, this mechanism is only possible if we identify the $U(1)$ sector with the hypercharge sector $U(1)_{Y}$.  In general the $U(1)$ could be a dark gauge field and in this case it would not couple to visible fermions and no leptogenesis will be possible with the mechanism that we will proceed to outline.  
Furthermore, the leptogenesis mechanism, while compelling, will need additional ingredients, such as a rapid reheating mechanism before we can claim with more confidence that leptogenesis actually occurs.  In what follows we present encouraging clues that the Sakharov conditions \cite{Sakharov} occur quite naturally during inflation.  We will explicitly calculate the net lepton number produced by the gauge fields during inflation, using the one-loop ABJ anomaly \cite{ABJ}, allowing us to constrain the initial amplitudes $A^0_-$ and $A^0_+$ that are necessary to drive inflation.  In what follows, we show how each of the Sakharov conditions are satisfied in our model. 

Primordial nucleosynthesis and the recent determination of the cosmological parameters from the cosmic microwave background observations by the WMAP satellite require an excess of baryon to entropy density ratio \cite{WMAP} in the universe to be
\be\label{baryond}
\frac{n_B}{s} = (6.5\pm 0.4)\times 10^{-10}\ ,
\ee
where $n_B= n_b- n_{\bar b}$ and $s$ is entropy density of radiation.  

\subsection{CP Violation} 

\noindent  It was shown by \cite{Alexander:2004us} that inflation provides a natural arena for satisfying the three Sakharov conditions for baryogenesis\footnote{For more details on the role of the Chern-Simons term in cosmology we refer the reader to \cite{LWK,SA,SG}.}.  In \cite{Alexander:2004us}, the lepton asymmetry was generated by CP violating gravitational waves during inflation.  In this model, however, it is the gauge field rather than gravitational waves that is responsible for the lepton asymmetry.  Specifically, the interaction between the pseudoscalar and Chern-Simons term in (\ref{perti}) sources CP asymmetric gauge field configurations.  The pseudoscalar coupling creates left and right asymmetry in the circular polarized gauge fields, $A_{+},A_{-}$.  The gauge fields that are generated no longer have definite transformations under CP:
\be 
[\hat{CP}]\vec{A}_{+} \neq \vec{A}_{-}\,. 
\ee
This condition can explicitly be checked by noticing that the left and right-handed gauge fields (\ref{evoluzione}) and (\ref{Aminus}) have unequal amplitudes.  A parity transformation exchanges the handedness of a gauge field plane wave but does not transform the amplitude.  Note that this form of CP violation is not explicit but dynamical since the birefringent amplitudes are sourced by the coherent evolution of the pseudoscalar, $\theta$, during inflation, which arises from solutions of $\theta$ presented in Sec.~\ref{consistent}.
 
\subsection{Lepton Number Violation}
 
\noindent It is well known that in the standard model, $B-L_{i}=0$, is conserved (where $i$ is the lepton flavor index).  However in the presence of a hpercharge axion coupling to the hypercharge Chern-Simons term, it is possible to generate net right-handed electrons from the presence of hypermagnetic fields \cite{Giovannini:1997eg,Joyce:1997uy,Rubakov:1986am}.  
According to those authors, the presence of hypermagnetic fields can get interconverted to a net right-handed lepton number through the hypercharge anomaly at energies above the electroweak scale.   In our model, non-vanishing hyper magnetic fields are indeed generated and will lead to a non-vanishing Chern-Simons number during inflation: 

\be \label{chiral}
\nabla_{\mu} \mathcal{J}^{\mu}_{e_{R}}= 
y_{R}Y_{\alpha \beta} Y_{\mu \nu} \epsilon^{\alpha \beta \mu \nu}/(32 \pi^2)\,,
\ee
where $Y_{\mu\nu}$ is the hypercharge field strength, $y_{R}=-2$ is the right handed electron hypercharge and $\cal{J}\rm^{\mu}_{e_{R}}$ is the right-handed electron current.

Specifically, during inflation, we demonstrated that the backreacted hypercharge field configuration exists, (\ref{evoluzione})-(\ref{Aminus}), and we will now proceed to show that it generates a non-vanishing  $Y_{\alpha \beta} Y_{\mu \nu} \epsilon^{\alpha \beta \mu \nu}/(32 \pi^2)$, resulting in a net right-handed electron number. 
During the Electroweak epoch $T \sim T_{EW}$, baryon plus lepton, $B+L$, violating sphaleron transitions will convert the net lepton number into baryon number via the relation: $n_{B}= \frac{4}{11}n_{L}$ \cite{Kuzmin:1985mm}.  We will pursue a more detailed account of the particle physics aspects of this leptogenesis mechanism in the context of preheating, which will require some numerical analysis \cite{SD}.

We now show that net lepton number resulting from eq (\ref{chiral}) is non-vanishing precisely because the gauge field fluctuations that occur during inflation is also parity violating, $\delta A(\eta, k)_{+}\neq \delta A(\eta, k)_{-}$.  This relationship between two Sakharov conditions is unique to our model. We can immediately calculate the amount of matter asymmetry produced from the gauge fields that initiated inflation through eq. (\ref{chiral}).   Through the triangle interaction the gauge-field converts itself into a net right-handed electron number accumulated throughout the inflationary epoch, and (\ref{chiral}) becomes:
\ba \label{bary}
n_{e_{R}} = \int_0^{H^{-1}} \!\!\!\frac{d\eta}{a^3(\eta)} \,k\,( \delta \dot{A}(\eta,k)_{+} \delta A(\eta,-k)_{+}-\delta \dot{A}(\eta,k)_{-} \delta A(\eta,-k)_{-})\simeq 
\frac{k\, \dot{\theta}}{ M_* H}
\, A^0_-\,A^0_+ \,,
\ea
in which we have used (\ref{evoluzione}) and (\ref{Aminus}), assumed for simplicity that $\tilde{A}^0_-=\tilde{A}^0_+=0$ and considered modes for which $k <\!\!< \dot{\theta}/M_*$ \footnote{Modes for which $k >\!\!>\dot{\theta}/M_*$ will be sub-dominant.}.  The non-vanishing Chern-Simons term arises from the perturbations of spatial components of the gauge field.  If we choose $\dot{\theta}/M_*\simeq 10^{-5} M_p$, and $A^0_- \, A^0_+ \simeq 10^{-10} M_p^2$, we find that the net lepton density,  $n\!\sim\!10^{-15} M_p^3$.  These values are all consistent with the conditions for inflation, namely $A_0 J^0\sim 10^{-10} M_p^4$, and with the inequalities required by isotropy, $|A^0_- |<\!\!< A_0$ and $|A^0_+|<\!\!< A_0$.  Therefore, for modes that extend up to horizon, (\ref{bary}) gives as result $n\!\sim\!10^{-15} M_p^3$ whenever $A^0_- \, A^0_+ \dot{\theta}/M_*\simeq 10^{-15} M_p^3$.

We can use the Friedmann equation to find the entropy density, $s=1.8 g_* n_{\gamma}$, where $g_*$ is the effective number of massless degrees of freedom ({\it e.g.} $100$ for the MSSM) and $n_\gamma\!=\! 1.28 g_*^{-3/4}(HM_p)^{3/2}$.  Hence for modes extending up to the horizon, we calculate a realistic value for the baryon asymmetry index $n/s\sim 10^{-10}$.   Note that the time evolution of the gauge field, $\delta {A}_\pm$ is crucial in order to obtain the correct lepton-number, which shows the interesting relation between the dynamics of the gauge field during inflation and leptogenesis.  We intend to pursue the real time dynamics of the lepton number production in a forthcoming paper.

\subsection{Out of Equilibrium}

\noindent Lepton number and CP violation occur simultaneously during inflation through the gauge-fermion vertex and triangle loop diagram respectively. However, the exponentially expanding background space-time renders the universe to be far out of equilibrium with the fermion production.
Furthermore, because we are generating the Sakharov conditions during inflation, this leptogenesis mechanism depends on the details of a reheating mechanism to conserve the net-leptons produced during inflation.  We assumed a spontaneous reheating \cite{Linde}. However if reheating occurs slowly to a reheating temperature $T_{r} < T$, then the baryon asymmetry index, ${n}/{s}$ will be diluted by a factor $T_{r}/T$.   
Therefore, to firmly establish that we have a successful baryogenesis mechanism will require a detailed analysis of reheating in this model; a topic that we leave for future work.

\section{End of Inflation and Scale-Invariant Density Perturbations}

\noindent In most scalar models of inflation, inflation ends when the inflaton no longer satisfies the slow-roll conditions. Our model is similar in that the end of inflation is triggered by the late time oscillation of the $\theta$ field about its minimum.  Recall that inflation is driven by a nearly constant interaction energy between the gauge-field and the fermionic current, which dominates the energy momentum tensor. The solutions of the gauge field's (background and perturbation) components, which we have found in Sec. \ref{gauge} and are responsible for inflation, arose from a phase where the velocity of the $\theta$ field is slowly-rolling ({\it i.e.} $\dot{\theta} \sim {\rm const})$:
\be 
\delta \ddot{A}(\eta,k)_h + k^2 \delta A(\eta,k)_h +h \, k\, \delta A(\eta,k)_h \dot{\theta}/M_* = 0 \,.
\ee
Eventually, the $\theta$ field, which couples to the vector field, evolves to the bottom of the its potential, begins to oscillate, and the slow-roll condition is violated, $\epsilon =\frac{M_{p}^{2}}{2} \left(\frac{V'(\theta)}{V(\theta)}\right)^{2} \geq 1$.  

Even though the scalar potential does not source inflation, its eventual oscillatory coupling to the gauge field will end inflation.  Hence, following the work of Kofman et al. and Traschen et. al \cite{Linde2,Traschen}, at late times the scalar field approaches sinusoidal behavior, 
\be \label{sol}
\theta = \frac{M_{p}}{3\pi m \eta} \sin(m\eta) \,,
\ee
($m$ being the mass of the inflaton) resulting in a Mathieu equation which yields a resonant particle production:
\be \label{mat}
\delta \ddot{A}(\eta,k)_h + k^2 \, \delta A(\eta,k)_h + \Phi_{0} \sin(m \eta) \, \delta A(\eta,k)_h=0 \,,
\ee
where the amplitude is $\Phi_{0} = \frac{k M_{p}}{3\pi M_* \eta} $. Notice that we can use (\ref{sol}) because of the fast roll condition ($\epsilon>1$), as the kinetic term  $\ddot{\theta}$ dominates over Chern-Simons potential.

As a result of the oscillation of the $\theta$ field about its minimum, fluctuations of the gauge field cease to propagate as waves and start growing exponentially, which will spoil the isotropy conditions necessary to maintain inflation. This happens because the Chern-Simons coupling of $\theta$ to the gauge field mimics a Mathieu equation (\ref{mat}) for the gauge field. This leads to resonant particle production similar to what happens in parametric resonance in preheating.  Since the equations are non-linear we leave this preheating analysis up to future numerical investigations.  

Similar to scalar field driven inflation, the scale invariant density perturbations will be generated by the quantum fluctuations of the pseudo scalar field $\delta\theta(x,t)$ since this field slowly rolls during inflation, obeying the familiar equation
\be \label{M}
\delta\ddot{\theta}+3H \delta\dot{\theta} + 2 m^2 \delta\theta =  \frac{\vec{E}\!\cdot\!\vec{B}}{4 a^3 M_*}\,.
\ee
According to the analysis of \cite{AS}, the scalar mode functions will generate a scale invariant spectrum provided that the r.h.s. does not generate appreciable back reaction.  In the following section we show that back reaction is indeed negligible.

\section{Consistency of the Analysis} \label{consistent}

\noindent We wish to address the evolution and issue of backreaction on the $\theta$ field due to the Chern-Simons term. Eventually, the Chern-Simons term could provide a back-reaction potential for $\theta$ that would be responsible for the non-constancy of $\dot{\theta}/M_*$. However, if we set up sufficiently small values of the initial conditions for the perturbations of the vector field, we derive a disregardable maximum value (at the beginning of inflation) for the back-reaction due to the Chern-Simons term. Such a value will be then suppressed by a factor $e^{-240}$ at $\eta_f$, since $F_{\alpha \beta}\tilde{F}^{\alpha \beta}|_{\eta_0}\!=\! \vec{E}\!\cdot\!\vec{B}|_{\eta_0}/a(\eta_0)^4 \!\sim \! \beta_k\,\gamma_k\, 10^{-10}M_p^2$. Thus the Chern-Simons term would provide a negligible back-reaction to $\theta$. It is straightforward to check this statement by working in co-moving coordinates and then pull-back the result in conformal coordinates using $d\theta/d\eta=a(t)d\theta/dt$. In what follows we focus on modes extending up to the horizon and set $k\simeq H$, which reasonably fixes the highest initial bound for the Chern-Simons term to be $F_{\alpha \beta}\tilde{F}^{\alpha \beta}|_{\eta_0}\! \sim \! 10^{-20}M_p^4$.

From (\ref{action1}) the time evolution of $\theta$ in a homogenous FLRW metric with $a\!=\!a_0e^{Ht}$ is determined by
\be \label{M}
\ddot{\theta}+3H \dot{\theta} + 2 m^2 \theta =  \frac{\vec{E}\!\cdot\!\vec{B}}{4 a^3 M_*}\,.
\ee
We first study the general solution of (\ref{M}) ignoring the Chern-Simons term. The conformal coupling with the metric provides a Hubble friction term $3H\dot{\theta}$ responsible for a dramatic decay of $\dot{\theta}/M_*$ in both conformal and co-moving time.  However, the potential acts to prevent the decay in conformal time of $\theta$ during inflation, provided that $m=H$, namely $m$ is order GUT scale. With an appropriate choice of the initial condition, the general solution in co-moving coordinates is $\theta\!=-\theta_0\, e^{-H t}$. In conformal coordinates, such a solution would give the constant value $d\theta/d\eta=\theta_0 H$. As we have set up $H\simeq M_*$,  the requirement in conformal coordinates $\dot{\theta}/M_*\!\simeq \! M_* \simeq \! H$ is satisfied provided that $\theta_0\!\simeq\!10^{-5}M_p$. Now we solve for $\theta$ with the Chern-Simons term present, which slowly varies during inflation. Therefore, the particular solution to (\ref{M}) becomes  
\be
\theta_p= \frac{(\vec{E}\! \cdot\! \vec{B})_0}{4 \,a_0^3 (2m^2-9H^2)  \,M_*} e^{-3H t}\,,
\ee
which at initial time of inflation contributes to the value of $\dot{\theta}/M_*$ with a term $\dot{\theta}_p/M_*=10^{-20}M_p$, after washed out by a factor $a^2(\eta)$. And the contribution to the minimum of the field from $\theta_p$ will be $\theta_p(0)/M_*=10^{-15}$, thus negligible with respect to the general solution to (\ref{M}). Therefore, at initial time we find the ratio 
\be
\frac{V(\theta_0)}{(A\!\cdot\! \mathcal{J})_0}\!=\!\frac{ m^2 \theta_0^{\,2}}{(A\!\cdot\! \mathcal{J})_0}\!\simeq\!  10^{-10},
\ee
that is exponentially suppressed at later times. It follows that potential scalar field $\theta$ does not directly contribute to driving inflation. 

If we do not fix the mass of the scalar field to be $m=H$, the Chern-Simons term may eventually provide a back-reaction that induces parity-violating CMB correlation functions among TB and EB modes \cite{Sorbo:2011rz}. In the limit $m/H<\!\!<1$, we approach for the scalar field the conformal-time behavior $\dot{\theta} \propto a(\eta)$, which under a proper choice of the initial conditions would lead to eq. (\ref{ser}), describing fluctuations of the gauge fields.

\section{Conclusion and Discussion.}

\noindent For many years, models of cosmic inflation have been pursued using fields that exist beyond the standard model.  In this work we demonstrate that it is possible to obtain an epoch of cosmic inflation from fields that already exist in nature, specifically the time-like $U(1)$ gauge field interacting with a fermionic charge density.  This interaction leads to an exponential time dependence of the gauge field that compensates the diluting fermionic charge to give a nearly constant energy density $\rho_{AJ} \sim A\!\cdot\! \mathcal{J}$.  

Furthermore, we related and constrained the amplitudes of the spatial gauge field fluctuations to the observed the baryon asymmetry index; these gauge field fluctuations are inevitably generated during inflation.  However, we assumed that the entropy production occurs from spontaneous reheating.  We did not pursue a reheating mechanism inherent to this model.  Although, we argued that parametric particle production of the gauge particles due to coherent oscillations of the $\theta$ field triggers the end of inflation, which could yield a preheating phase, and we shall pursue this analysis for future work. 

We showed that scalar field is already sufficient to reproduce the nearly scale-invariant powerspectrum observed in CMBR provided that back reaction of the Chern-Simons term is negligible.  It might be tempting argue that an abelian vector field, which in this work has been shown to successfully drive inflation, could also be used in explaining the CMBR spectrum.  An explicit analysis of the generation of scale invariant spectrum of density perturbation has been initiated in a parallel project \cite{ASM2}, in which we have found that both tensor and vector metric perturbations do not diverge.  A more detailed perturbation analysis is necessary and is complicated by the fact that in the presence of spatial components of a vector field, the decomposition theorem is violated.  In particular, metric vector perturbations show up in the scalar perturbation equations \cite{ArmendarizPicon:2004pm}.  However, this coupling may lead to observable non-Gaussianity effects in the CMBR that will be analyzed in a future work \cite{ASM2}.  In closing, we believe that this model of inflation may play a role in connecting cosmological observables to particle physics in new ways, such as the baryon asymmetry index with the measure of anisotropy and non-Gaussianity in the CMBR.

\section*{ Acknowledgements} 
\noindent We would like to thank the referee for their useful and insightful comments, Robert Brandenberger, Robert Caldwell, Larry Ford, S. James Gates Jr., Alan Guth, Elias Kiritsis, Justin Khoury, Matthew Kleban, Paul Langacker, Alessio Notari, Mikhail Shaposhnikov, Lorenzo Sorbo, Herman Verlinde, Filippo Vernizzi and Edward Witten for enlightening discussions.

\end{document}